\documentclass[manuscript,12pt,anonymous=false,nonacm]{acmart}
\AtBeginDocument{%
  \providecommand\BibTeX{{%
    \normalfont B\kern-0.5em{\scshape i\kern-0.25em b}\kern-0.8em\TeX}}}

\setcopyright{acmcopyright}
\copyrightyear{2020}
\acmYear{2020}
\acmDOI{10.1145/1122445.1122456}

\acmConference[FAccT '21]{FAccT '21: ACM Conference on Fairness, Accountability, and Transparency}{March 2021}{Virtual}
\acmBooktitle{FAccT '21: ACM Conference on Fairness, Accountability, and Transparency,
  March 2021, Virtual}
\acmPrice{15.00}
\acmISBN{978-1-4503-XXXX-X/18/06}

\usepackage[font=small,labelfont=sc]{caption}
\usepackage{subcaption}
\usepackage{url}

\usepackage{enumitem}
\setlist[enumerate]{topsep=0pt,itemsep=-1ex,partopsep=1ex,parsep=1ex}

\setcopyright{none} 
\renewcommand\footnotetextcopyrightpermission[1]{} 

\setcitestyle{round}
\usepackage{todonotes}
\usepackage{mathtools}

\begin{document}

\title{Authenticity and exclusion:‌ social media algorithms and the dynamics of belonging in epistemic communities}

\author{Nil-Jana Akpinar}
\affiliation{%
  \institution{Amazon AWS AI/ML}
  \city{Seattle}
   \country{United States}
}
\email{nakpinar@amazon.com}
\authornote{Both authors contributed equally to the paper}
\authornote{Work done outside of Amazon.}

\author{Sina Fazelpour}
\affiliation{%
  \institution{Northeastern University}
  \city{Boston}
   \country{United States}
}
\email{s.fazel-pour@northeastern.edu}
\authornotemark[1]

\begin{abstract}
Recent philosophical work has explored how the social identity of knowers influences how their contributions are received, assessed, and credited. However, a critical gap remains regarding the role of technology in mediating and enabling communication within today's epistemic communities. This paper addresses this gap by examining how social media platforms and their recommendation algorithms shape the professional visibility and opportunities of researchers from minority groups. Using agent-based simulations, we investigate this question with respect to components of a widely used recommendation algorithm, and uncover three key patterns: First, these algorithms disproportionately harm the professional visibility of researchers from minority groups, creating systemic patterns of exclusion. Second, within these minority groups, the algorithms result in greater visibility for users who more closely resemble the majority group, incentivizing assimilation at the cost of professional invisibility.  Third, even for topics that strongly align with minority identities, content created by minority researchers is less visible to the majority than similar content produced by majority users. Importantly, these patterns emerge, even though individual engagement with professional content is independent of group identity. These findings have significant implications for philosophical discussions on epistemic injustice and exclusion, and for policy proposals aimed at addressing these harms. More broadly, they call for a closer examination of the pervasive, but often neglected role of AI and data-driven technologies in shaping today's epistemic communities. 
\end{abstract}

\maketitle
\pagestyle{plain}

\section{Introduction}\label{sec:intro}
Imagine you're a minority researcher sharing your work on social media, only to see it vanish while similar posts from colleagues gain traction. Is this just bad luck, or something more systematic? Recent works in philosophy have explored such questions, showing how scientists' social identities can influence their network and reach, and the way their work is received, assessed, and credited~\cite{rubin2022structural,fazelpour2022diversitysteel,o2019origins,dotson2014conceptualizing,wu2023epistemic,davis2018epistemic,settles2021epistemic}. While these studies illuminate key ethical and epistemic implications of identity-based dynamics for epistemic communities, they often overlook the critical role of \textit{technology} in mediating and enabling communication within these communities. This paper fills that gap by examining how social media platforms and the recommendation algorithms that are at their core shape the visibility and professional opportunities of users from minority backgrounds.

Social media platforms have become indispensable tools for many academics~\citep{donelan2016social}. Researchers use these platforms to share their work, engage in scholarly discourse, seek and advertise opportunities, forge collaborative ties, and build professional networks~\citep{britton2019reward,hurrle2015social,luo2020like,knight2016tweet,klar2020using,veletsianos2016social,neal2012social,sugimoto2017scholarly}. Many view these platforms as democratizing forces that make epistemic communities more accessible and inclusive~\citep{bartling2014opening}. Some even argue that alternative metrics derived from social media engagement offer a more accurate and comprehensive measure of scholarly impact than traditional metrics, and advocate for their use in academic promotion and tenure decisions~\citep{cabrera2018social,piwowar2013value}.

But does social media benefit all scholars equally? We argue that it can \textit{worsen} existing patterns of exclusion. To illustrate this, consider how \textit{homophily}---the attraction to similars---can shape communication, even in traditional settings. In diverse teams, members tend to share a joint professional interest, but their preferences for engaging in other topics often align with their identities (e.g., cultural products or activities). As a result, even though group identity does not influence discussions of professional topics of shared interest, it can drive engagement when conversation shift to other topics, leading to preferential engagement with similar others. Research shows that homophily-driven dynamics create a tension between \textit{authenticity} and \textit{inclusion} for individuals from minority backgrounds~\citep{phillips2009,phillips2018diversity}. If they disclose their genuine interests, they risk exclusion from the broader conversation. Conversely, to gain inclusion, they might feel pressured to assimilate, suppressing their preferences. 

Importantly, the design and nature of social media platforms and their recommendation algorithms can aggravate these issues. On social media, \textit{context collapse}---the blurring of personal and professional spheres---is more pronounced and pervasive than in traditional settings, where such overlap is typically limited~\citep{boyd2010social,sugimoto2015friend,vitak2012impact}. What is more, recommendation algorithms that critically curate content visibility on these platforms can transfer engagement patterns from one context to another. Specifically, these algorithms might perpetuate a cycle where a lack of engagement in topics of differential interest can lead to decreased visibility of future posts, \textit{even on topics of shared interest} (e.g., research). This contrasts with traditional settings, where shared professional interests typically guarantees ongoing cross-group engagement on those topics of shared interest. In Section~\ref{sec:background}, we provide an overview of existing research on homophily, context collapse, and algorithmic bias in recommendation systems that informs our work. 

To explore how social media content recommendation algorithms interact with homophily-based communication, we use agent-based simulations. Our setup, which we detail in Section~\ref{sec:methods}, is inspired by components of Twitter's recently open-sourced, recommender system that, to the best of our knowledge, still has variations in production today~\cite{RealGraph2014,twittergithub}. In Section~\ref{sec:results}, we present three sets of key findings from our simulations. First, we observe a decrease in the visibility of professional content for minority groups, a disadvantage that is amplified by the recommendation algorithm over time. This means these algorithms create patterns of amplification and suppression that disproportionately harm minority users. Second, within minority communities, users who more closely resemble the majority group 
gain increased cross-group visibility for their professional content. In this way, the algorithms incentivize assimilation at the cost of professional invisibility. Finally, even for topics that strongly align with minority identities, content created by minority users is less visible to the majority than similar content produced by majority users. This raises concerns about recognition and representation.

Our findings extend philosophical work on understanding the systemic causes of harms against individuals and collectives in epistemic communities, especially in our algorithmically-mediated societies. This expanded understanding, in turns, enables a more thorough evaluation of proposals about the use of metrics that draw on social media engagements, and more generally algorithmically-mediated outcomes, in evaluating scientific merit and impacts. In Section~\ref{sec:discussion}, in addition to discussing these implications of our findings, we explore potential individual, technical, and policy-based interventions for counteracting these exclusionary patterns, and discuss future directions.

\section{Background and related work}\label{sec:background}
\subsection{Homophily and inter-group dynamics}
Homophily influences a wide array of social dynamics, including the formation of interpersonal associations, communication patterns, and the establishment and calibration of trust~\citep{golub2012homophily,o2019origins,McPherson2001}. This phenomenon is observed both offline and on digital platforms, where various salient attributes such as gender, country of origin, values, occupation, and more, can serve as axes of similarity~\citep{lawrence2020homophily,currarini2009economic,rubin2022structural}. In academia, for example, homophily (e.g., based on gender, race, or national origin) has been shown to influence interaction patterns at events such as conferences~\citep{atzmueller2018homophily}, patterns of citations~\citep{zhou2024gender}, even extending to professional opportunities such as prize nomination and selections~\cite{gallotti2019effects}. 


Homophily's influence on the dynamics of exchange and self-disclosure can be particularly burdensome for professionals from minority backgrounds~\citep{phillips2018diversity,ertug2022does}. Here, even if aligned in professional interests, when it comes to other topics, individuals may engage more readily in conversations with those with similar identities. Conversely, those concerned with a lack of engagement may be more reluctant to disclose aspects of themselves to others seen as out-groups~\cite{phillips2009}. As noted by~\citet{phillips2018diversity}, this reluctance to open up about aspects of oneself (other than those seen as aligned with shared professional interests) can have adverse consequences for one's professional success (e.g., promotion) and sense of belonging. While \citet{phillips2018diversity} focus on the impact of homophily on \textit{disclosure}, we examine how homophily's influence on the \textit{uptake} of information in social media can create \textit{incentives for assimilation} or lack of authentic disclosure. This happens because of the context collapse on social media platforms as well as the content recommendation algorithms that are at the heart of these platforms. 

\subsection{Context collapse and risks to minority and marginalized identities}
When using social media platforms, individuals, including academics~\citep{sugimoto2015friend}, must contend with how the platforms can collapse together different spheres of one's personal and professional life~\citep{davis2014context,boyd2010social}. It requires users to negotiate, across varied norms and expectations, what to disclose about themselves, when, and how~\citep{ellison2015use,hollenbaugh2021self}. 

Particularly for minorities, this merging of the (real and imagined) audiences and associated topics complicates the dynamics of self-presentation to professional audiences~\citep{vitak2012impact}. For many such users, social media platforms offer key spaces for joint identity construction with others with similar backgrounds, lived experiences, and interests~\citep{brock2009you,jackson2020hashtagactivism,khazraee2018digitally}. 
Importantly, the success of such projects can critically depend on keeping contexts, audiences, and expectations distinct (e.g., to create a safe space for disclosure; to ensure anonymity). 
The loss of distinctness can thus pose consequential privacy risks to these individuals~\citep{vitak2012impact,dhoest2016navigating}. In response, individuals can use various strategies to mitigate the risks or else the tensions due to context collapse. For example, they might sever or limit access (via ``unfriending'', ``blocking'', or ``muting'')~\citep{zhu2021context}.

While most previous works explore the risks due to the loss of privacy and anonymity, we examine harms that arise as a result of the interaction between homophily-based dynamics and content recommendation algorithms~\citep{phillips2018diversity}. Importantly, insofar as disclosing information can further career opportunities (e.g., by deepening interpersonal ties) in these professional settings, \textit{not} revealing information (whether by blocking or limiting access) will not mitigate the risks of homophily-based disparities.

\subsection{Bias and recommender systems}
Recommender systems are integral to curating user experience on social media platforms, shaping who and what people may (or may not) interact with on these platforms. Previous work has highlighted how these algorithms can perpetuate, exacerbate, or even generate biases that harm individuals, groups (especially those with minoritized and marginalized identities), and communities \cite{BaezaYates2018,pmlr-v81-ekstrand18b}. 
This issue has garnered attention in the field of algorithmic fairness, which has proposed numerous approaches to define, measure, and address unfairness in ranking and recommendation systems \cite{Akpinar2022,burke2017multisided,Zehlike2022,Geyik2019,Akpinar2022leqi,Celis2018,Biega2018,Amig2023}.\footnote{We refer to the works of \citet{Patro2022} and \citet{Chen2023} for a synthesis of the current state of the fair recommendation area.}


Research on ``algorithmic glass ceilings'' in recommendation systems has revealed how biases related to gender and social similarities are perpetuated, impacting fair representation. Such systemic barriers, prevalent in various platforms, hinder the visibility of women and minorities, and have concerning fairness implications~\cite{Stoica2018,Avin2015}. A different line of work describes the ``filter bubble problem of link prediction'' \cite{Masrour2020,Nguyen2014}.
Homophilic tendencies in network formation and demographic disparities in connection recommendation exacerbate the isolation of minority groups, limiting their visibility as well as their access to diverse perspectives \cite{Hofstra2017,Fabbri2020}. Research like the work of \citet{Akpinar2022} demonstrates that, while short-term fairness interventions in social connection recommenders might initially appear effective, they fail to address long-term bias amplification.

Agent-based modeling (ABM) has emerged as a prominent method for understanding the longitudinal dynamics of algorithms and interventions on social media~\cite{adomavicius2021understanding,Patro2022,Akpinar2022,Chaney2018,mansoury2020feedback}. ABM describes simulation approaches in which heterogeneous \textit{agents} (e.g., users) interact with each other in a controlled \textit{environment} (e.g., recommendations in a social network graph). Although modeling takes place at the individual level, the underlying goal is typically to extract insights about emerging phenomena at the level of the collective over longer runs.\footnote{For this reason, agent-based modeling has also emerged as a valuable tool for understanding social dynamics in works in social epistemology, philosophy of science, and political philosophy~\citep{zollman2010epistemic,vsevselja2022agent,o2019origins,muldoon2013diversity}.} ABM can provide a particularly useful tool, when real-world experimental (e.g., A/B testing) and offline (e.g. static data) recommendation settings are limited to the study of short-term effects due to financial and ethical concerns or simply because controlling for confounding factors and spurious effects over time is difficult, if not impossible \cite{krauth2020offline}. For these reasons, agent-based simulation has been used in the recommender systems context to study homogenization effects \cite{Chaney2018}, filter bubbles \cite{Aridor2020}, performance paradoxes \cite{adomavicius2021understanding}, reinforcement learning for search ranking \cite{hu2018reinforcement}, popularity bias in search engines \cite{Fortunato2006}, effects of fairness intervention \cite{Akpinar2022}, and more \cite{Patro2022}.

\section{Methods}\label{sec:methods}

Social media users engage in communication by creating and sharing content, and interacting with others' content (e.g., through comments, likes, and shares). What these users communicate, and whether and how they interact with the content they are presented with are shaped by their interests, aims, and preferences, which are in turn influenced by their cultural backgrounds and identities. But \textit{what} they are presented with and \textit{whether} they see some content at all are inherently mediated by recommendation algorithms, whose objective is to maximize user engagement. These algorithms function as hidden curators of the digital world, shaping users' experiences and the public discourse online, by amplifying certain voices or topics, while deprioritizing others~\cite{gillespie2010politics}. In this section, we describe how we model this complex sociotechnical system in our simulations.

\subsection{Users, topics, and preferences}
In our simulation, we categorize users into one of two groups: a majority group $G_0$ and a minority group $G_1$. The proportion of users in each group is given by $p_0$ and $p_1=1-p_0$ respectively. To capture the idea that social media enables users to engage in conversations reflecting different aspects of their identities and interests, we consider three distinct topic groups: 
\begin{enumerate}
    \item \textbf{Professional topics}: Work-related topics that are of equal relevance to both groups.
    \item \textbf{Mainstream topics}: Topics of high interest to the majority group but less so for the minority group.
    \item \textbf{Marginal topics}: Topics of high interest to the minority group but less so for the majority group.
\end{enumerate}
In academic settings that are our focus here, for instance, users from different groups might have a shared interest in content related to certain research articles, ideas, and professional news and opportunities (professional topic). But, when the conversation shifts to other topics, their interests may diverge, clustering around a mainstream and a marginal topic in ways that are correlated with their group identity. This formulation is intentionally vague to encompass a wide array of scenarios that to our understanding are compatible with the workings of the recommendation algorithms described below. For example, it allows for the mainstream and marginal topics to be about different cultural products (e.g., TV shows from different cultures, different genre of music, ...), activities (e.g., running, rock climbing, yoga, ...), and more. And it even allows these topics to be about different \textit{sub-disciplines} in a field, or different disciplines in an inter-disciplinary field.

We model each user as having a set of preferences that guide their interest in communication about these different topics. For simplicity, we assume that these preferences reflect both sides of communication around that topic---that is, users' likelihood of \textit{posting} about the topic as well as their likelihood of \textit{interacting with} others' posts on that topic. Formally, we model the preference of users in a topic category as random draws from a group-dependent normal distribution $z^\text{topic}\sim\mathcal{N}(\mu^\text{topic}_G,\sigma^\text{topic}_G)$. The preference vectors of users are, then, their set of preferences about professional, mainstream, and marginal topics, with the $i$-th entry in $z$ encodes user's interest in topic $i$. 

Using different average values for preferences about topics from each group allows us to model how users' identities shape their engagement with topics, thus representing the effects of homophily on communication dynamics. In our simulations, in particular, we used $\mu_{G_1} = (0.5, 0.1, 0.4)$ for the average interests of the minority group in the above topics, and $\mu_{G_0} = (0.5, 0.4, 0.1)$ for that of the majority group. These values indicate that, on average, both groups have a strong interest in professional content ($0.5$), but differ in their interests in mainstream ($0.1$ for minority, $0.4$ for majority) and marginal ($0.4$ for minority, $0.1$ for majority) topics. We intentionally select mean topic preference vectors that demonstrate a pronounced collective interest in the professional topic, alongside reduced interests in the other two topics, which exhibit a strong correlation with group membership. 

We set the variance around these average to $\sigma^c_G=0.1$ for all topics for both groups. Importantly, the variation in individual preferences wallows for some diversity of interests within each group. This enables us to examine the potential effects of ``resemblance'' and ``assimilation'' (e.g., minority users whose interests are more similar to the majority group, compared to their other group members) on inter-group engagement and overall visibility. 


\subsection{Social networks}
Social media platforms afford connectivity by allowing users ``befriend'' or ``follow'' others. 
Here we focus on the follow relation, and model the resulting digital networks using \textit{directed} graphs. In particular, we consider the following graph structures:
 
\paragraph{Complete network} A fully connected graph, where every user follows every other users, and so is eligible to be recommended content by them. 

\paragraph{Random network} A graph where not all possible connections are realized, but rather every pair of users $i$ and $j$ has the same probability $p^{\text{edge}}$ to form an edge. Note that, in order to obtain a directed graph representing the ``follow'' relation on social media, an edge from $i$ to $j$ is sampled independently from a potential edge from $j$ to $i$. For the random graph structure, we assume a fixed directed edge probability of $0.5$.\footnote{Note that because of the imbalanced group shares this implies that the average majority group user has a following that consists of 79.98\% majority group users, and users of the minority group have a following that comprises 19.92\% minority group users.}

\paragraph{Homophilic networks} We also consider a homophilic network wherein users are more inclined to be connected with others within the same group rather than with those from different groups. As mentioned above, the presence of such homophilic tendencies in network formation is supported by social science literature \cite{Kossinets2006,Louch2000,McPherson2001}, and their implications are a burgeoning topic of philosophical examination~\cite{o2019origins,rubin2022structural,fazelpour2022diversityhomophily,lacroix2021dynamics}. Considering these networks is also important as they can be the natural results of many affinity groups, including those created by minority groups as sites for seeking social support or joint identity creation. We model connections in this type of graph using a stochastic block model with parameters $(p^{\text{edge}}_{\text{maj}}, p^{\text{edge}}_{\text{min}}, p^{\text{edge}}_{\text{cross}})$. Here, $p^{\text{edge}}_{\text{maj}}$ is the probability with which a user from the majority group follows another majority group user, $p^{\text{edge}}_{\text{min}}$ is the probability for a directed edge within the minority group, and $p^{\text{edge}}_{\text{cross}}$ is the probability for cross-group edges. In our simulations, we use the default parameters $p^{\text{edge}}_{\text{min}}=0.5$, $p^{\text{edge}}_{\text{maj}}=0.4$, and $p^{\text{edge}}_{\text{cross}}=0.1$. These default values are selected to produce a graph structure in which individuals are more likely to follow users of the same group, and the minority group has an increased interest in community forming.\footnote{On average, minority group users have 54.94\% minority group followers while majority group users have 94.32\% majority group followers with these parameters.} 

Of course, social media networks can change dynamically as a result of user actions, such as by following or unfollowing content creators, or blocking specific types of content. In order to exclusively examine the impact of recommendation algorithms, however, we keep the graph structures fixed. This means that, throughout the observation period, users within the graph do not establish new connections.

\subsection{Recommendation policy}
\label{sec:rec_policies}
Content recommendation by social media algorithms involves selecting and curating posts to display to users from the set of all candidate content that they could be shown. To formalize this task, let $C$ refer to the set of all candidate contents available for user $i$ at time $t$. Let us assume $C$ only includes all content created at time $t$ by users that $i$ follows. This is a simplifying assumption. For example, social media platforms consider as candidate sources of posts to display to users both posts from accounts that they follow (so-called ``In Network'' sources) \textit{and} posts from accounts that they do not currently follow (so-called ``Out-of-Network'' sources). The latter includes sponsored content, trending posts, or else content predicted to be engaging to users, even if they do not follow the content creator. We do not consider these Out-of-Network sources because, in addition to the added complexity, their selection is carried out by different types of algorithms than those used for In Network selection, which is our focus here. As we will discuss in Section~\ref{sec:discussion}, however, given the typical conceptualization of the task of those Out-of-Network content recommendation algorithms by social media companies, we might expect that they \textit{exacerbate} the problems that we identify in this paper.

We can formalize the task of content recommendation as assigning each content $c \in C$ a scored by a scoring function $s$, and then retrieving a recommendation by sampling from the normalized vector of scores, such that items with higher scores are more likely to get recommended. Accordingly, different recommendation policies amount to different scoring procedures. We consider the following procedures. 

\paragraph{Random} Each content item $c$ is assigned the same score $s=s(c)$, and so is equally likely to get recommended. Given our setup, this could be seen as a type of chronological timeline, in which all items created in time-step $t$ have priority over items created previously, but, compared to each other, are equally likely to get recommended. 

\paragraph{Topic match} The score of content item $c$ is given by the user's preference for the topic of $c$. That is, $s(c)=z^{\text{topic}(c)}$ where $z$ is the preference vector of user $i$. 

\paragraph{RealGraph} The content recommendation algorithms on social media typically depend on factors beyond content-level features (e.g., the topic of a single post). While there tends to be secrecy about how exactly this involves, Twitter's recently open-sourced algorithm offers important insights into key components of a highly influential recommendation system, and potentially those of other platforms. As \textit{Twitter} describes in a blog post, their recommendations depend on models that seek to predict key \textit{user-level} measures.\footnote{As mentioned above, here we are only focusing on recommendation from the pool of those followed by user (or In Network sources). We will discuss some implications of Twitter's recommendation algorithms for Out-of-Networks sources in Section~\ref{sec:discussion}.} 
Specifically, as they explain, the ``most important component in ranking In-Network Tweets is Real Graph ... a model which predicts the likelihood of engagement between two users''~\cite{twitter}. The recommendation policy we describe below is directly inspired by the RealGraph procedure, whose details can be found in \citet{RealGraph2014}.

Besides content-level features, the RealGraph model involves a procedure for measuring the \textit{tie strength} between users, which relies on commonalities between features of the users (e.g., their social networks and preferences) as well as their interaction history. Following the descriptions in \citet{RealGraph2014}, we model this tie strength (or edge weight) procedure for estimating, $p_{ij}$, which represents the probability of interaction between user $i$ and user $j$, using a logistic function 
\begin{align}
\label{eq:tie_strength}
    \log \frac{p_{ij}}{1-p_{ij}}= \beta_1\ \text{outEdge}(i,j)+ \beta_2 \ \text{inEdge}(i,j) + \beta_3 \ \text{dist}(i,j)+ \beta_4 \ \text{intCount}(i,j)
\end{align}

Here, $\beta$ is a set of fixed parameters, $\text{outEdge}(i,j)$ and $\text{inEdge}(i,j)$ are indicators of similarity in social networks of $i$ and $j$, representing the number of common incoming (e.g., followers) and common outgoing (e.g., following) edges between the users respectively. $\text{dist}(i,j)$ is the Euclidean distance between users' topic preference vectors, 
offering another measure of (dis)similarity between the two users. 
%
$\text{intCount}(i,j)$ gives an indication of the number of times user $i$ has interacted with (e.g., commented on, liked or shared) content created by $j$ in the past. Following \citet{RealGraph2014}, we model $\text{intCount}(i,j)$ using an exponential moving average of their interactions that, while depending on all of $i$'s prior interactions with content by $j$, puts more weight on more recent interactions and exponentially less weight on older ones.\footnote{The use of exponential moving averages (or exponential smoothing) is also typical practice when dealing with noisy time series data~\cite{heckert2002handbook}.} Specifically, if at time $t>0$, content from $j$ was recommended to $i$, then the interaction count $\text{intCount}_t(i,j)$ is updated as
\begin{align}
\label{eq:int_count_mavg}
    \text{intCount}_t(i,j) = \alpha \ \text{intNumb}_t(i,j) + (1-\alpha) \ \text{intCount}_{t-1}(i,j)   
\end{align}

Where $\text{intNumb}_t(i,j)$ is the number of times $i$ interacted with content created by $j$ in time step $t$, $0<\alpha<1$ is a smoothing factor, and $intCount_0(i, j)=0$. If no content by $j$ was recommended to $i$, then the interaction count remains unchanged.\footnote{For our simulations, we use parameter vector $\beta = (1,1,-1,5)$ for Equation~\ref{eq:tie_strength}. This is informed by several assumptions. First, common followers and following accounts are associated to higher tie strength between two users. Second, the more similar---in terms of topic preferences---two users are, the stronger their tie. Since we measure distance between users (reverse similarity), the corresponding parameter is negative. We assume that the the common incoming and outgoing edges as well as the similarity of topic preferences have effects of similar magnitude.
Finally, we assume a large positive effect of smoothed interaction count on users' tie strength. This is reasonable because the interactions counts are the only features that vary over time and increasing their parameter in the tie strength model speeds up our simulation and allows us to observe meaningful results within 10,000 time steps.
For interaction count smoothing in Equation~\ref{eq:int_count_mavg}, we set $\alpha=0.01$. With increasing $\alpha$ the influence of previous interactions decays even faster. This could hurt minorities even more. In terms of our results, for example, this would mean that the decline in curves in Figure~\ref{fig:real_graph_inter_count_rec_0} will be even steeper.} 



The recommendation policy inspired by RealGraph, then, involves two steps that combine content-level and user-level factors: (i) retrieving two sets of scores via the \textit{topic match} procedure and the\textit{ tie strength} procedure just described, and (ii) using the average of the two scores. While our model is simplified, the use of exponentially decaying interaction counts, common follower and following counts, similarly in interests and logistic regression modeling is directly based on previous work on RealGraph~\cite{RealGraph2014}. Twitter's publicly available code library \cite{twittergithub} suggests that some alteration of the RealGraph framework may still be in production today. Yet we hypothesize that, while not revealed to the public, reliance on historical user-level interaction data as well as measures of commonality between users are commonplace among social media recommendation algorithms and keep our discussion on a general platform agnostic level.

\subsection{Simulation procedure}
Given this model, we can explore the effects of content recommendation policies on visibility in diverse epistemic communities by simulating recommendations in a network of $n$ users over $T$ time steps. 
At each time $t\in[T]$, our simulation goes through the following steps. 
\begin{enumerate}
    \item \textbf{Content creation step:} Every user in the network independently creates a new content item with probability $p_c$. The topics for created content are then sampled according to the creators' topic preference vectors, which as described above are shaped by their group membership. In our simulations, we set $p_c=0.2$. 
    \item \textbf{Recommendation step:} Users independently seek out content recommendations (e.g. by logging into the platform, or clicking on a particular tab on the website), with probability $p_r$. Only one piece of content is recommended at a time. Recommendations are retrieved according to one of the recommendation policies outlined in Section~\ref{sec:rec_policies}. We set $p_r=0.8$. 
    \item \textbf{Interaction step:} After recommendations are served, we sample whether users interact with or disregard the suggested content based on topic match. More concretely, a user with normalized preference vector $z$ interacts with recommended content $c$ with probability $z_{\text{topic}(c)}$.
    Note that interaction here is a binary concept and could refer to liking, commenting, or sharing the content. 
    \item \textbf{Update step:} Given the observed interactions, we first update the smoothed interactions counts from Equation~\ref{eq:int_count_mavg} and use it to update the tie strengths from Equation~\ref{eq:tie_strength}. All feature matrices are standardized before computing the updated tie strengths. 
\end{enumerate}
\section{Results}\label{sec:results}

Experiments are conducted for different variations of network structure, group sizes, and recommendation policies. For each simulation, we assume 1,000 users and simulate recommendation over 10,000 time steps. To allow the system adequate time to stabilize before making any assessments, we exclude the initial 2,500 time steps from our analysis. If not specified otherwise, we assume a 20\% / 80\% split of the user population into minority and majority groups. We conducted additional experiments to study the robustness of these findings with respect to parameter changes and find that the effects discussed below qualitatively persist when varying group sizes, as long as the minority group remains small; when adjusting the stochastic block model parameters; and when altering the parameters of the tie strength model, provided that prior interaction counts carry sufficient weight. Results are reported as averages over simulation runs.

\subsection{Professional content less promoted for minority group}
\label{sec:4.2}

\begin{figure}[ht]
    \centering
    \begin{subfigure}[b]{0.45\textwidth}
        \centering
        \includegraphics[width=\textwidth]{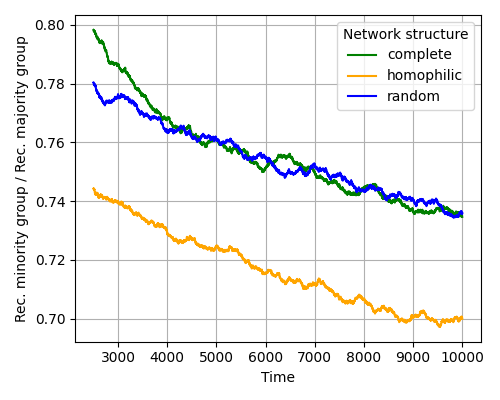}
        \caption{Moving average (window size = 1,000) of ratio of the average number of recommendations for professional topic content created by minority and majority group users over time.}
        \label{fig:real_graph_ratio_rec_0}
    \end{subfigure}
    \hfill
    \begin{subfigure}[b]{0.45\textwidth}
        \centering
        \includegraphics[width=\textwidth]{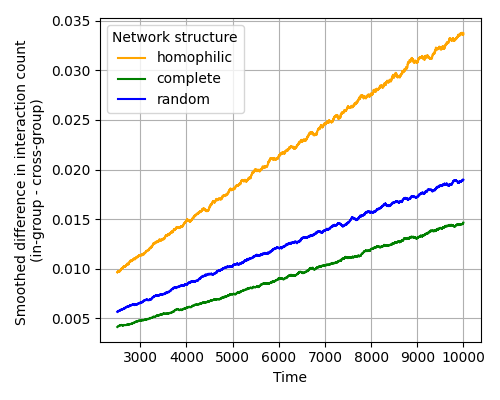}
        \caption{Moving average (window size = 100) of difference in interaction count of user pairs served professional topic recommendations over time.\\}
       \label{fig:real_graph_inter_count_rec_0}
    \end{subfigure}
    \caption{Recommendation results for professional topic content using \textbf{real graph recommendation policy with colors indicating different network structures}.
    Results are averaged over 20 simulation runs.
    }
    \label{fig:realgraph_topic0_plots}
\end{figure}

We first consider recommendations of professional topic content from content creators' point of view. Visibility of scholarly posts on online platforms can lead to significant professional advantages, such as generating productive discussions and feedback, increasing the reach of published research, and attracting citations. Additionally, it can offer social benefits, including networking with fellow scholars, gaining recognition within academic communities, and creating opportunities for collaboration or invitations to speak at conferences.


Figure~\ref{fig:realgraph_topic0_plots} depicts the professional content results of our experiment using the real graph recommendation policy for different graph structures. 
We see that professional topic posts created by minority group users in homophilic networks are, on average over users, time steps and simulation runs, recommended 72.36\% as often as comparable posts created by majority group members.
Furthermore, the recommendation policy perpetuates the disadvantage for the minority group over time. 
This amplification effect is explained by a feedback loop of recommendations served in one time step on the tie strength between users in the next step. The tie strength between users is a function of agreement between users' interests, the number of common `followers' and `following' users, and the smoothed number of previous interactions.
Real graph recommendation policy and variation in topic interests lead to interaction counts that are increasing faster for in-group pairings of users than cross-group pairings (Figure~\ref{fig:real_graph_inter_count_rec_0}) ultimately resulting in more recommendations of professional content for the majority group based on their size advantage.



Overall, our findings suggest that homophilic graph structure leads to decreased visibility of professional content for minority groups.
This disadvantage is perpetuated and amplified by recommendation algorithms like the real graph policy which are based on historical interaction counts.
Variations of the experiments with random and topic match recommendation policies reveal that, as long as a homophilic network is used, users are disadvantaged. However, the \textit{amplification} of this disadvantage over time does \textit{not} occur with these alternate recommendation policies. 

\subsection{More visibility for minority users who resemble the majority group}

\begin{figure*}[ht]
    \centering
    \includegraphics[scale=0.5]{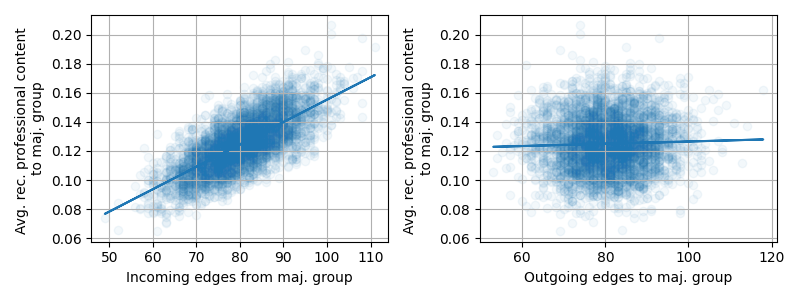}
    \includegraphics[scale=0.6,width=\textwidth]{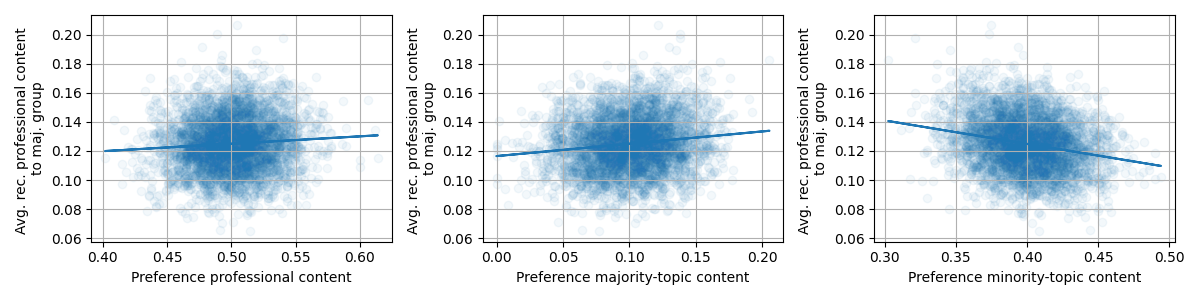}
    \caption{
    Average number of times professional content created by the minority group is recommended to majority group members in relationship to the number of incoming and outgoing edges and topic preferences. Each point corresponds to one minority group user in one 20 simulation runs using a \textbf{homophilic network structure and real graph recommendation policy}. Lines indicate linear approximations.}
    \label{fig:real_graph_homophilic_minority_work_rec_to_maj}
\end{figure*}

The most important driver for recommendations are follower counts: the more incoming edges (or followers) a user has, the more often we can expect their posts to be recommended to others.
Professional posts by minority group users are no exception to this. Yet stratifying our simulation results by group membership of followers reveals an interesting picture. While, on average over all minority group users, 45.02\% of followers belong to the majority group, only 15.07\% of professional post recommendations are made to the majority group.
Despite considerable followings from both groups, the professional content of many minority group users remains almost invisible to the majority group.  

We investigate the characteristics that influence successful recommendation of minority-created professional content to the majority group.
As expected, Figure~\ref{fig:real_graph_homophilic_minority_work_rec_to_maj} shows that minority group users with more followers from the majority group get their professional content recommended to the majority group more frequently with ($\rho=0.72$, $p\text{-value}<0.05$)\footnote{Pearson correlation coefficient $\rho$ and $p$-value of two-sided correlation $t$-test on data from 20 simulation runs.}. While this effect appears particularly strong, we observe various other interesting correlations.
Surprisingly, minority group users who follow a greater number of majority group members seem to gain a slight edge in receiving recommendations for their professional content on average ($\rho=0.04$, $p\text{-value}<0.05$). It's important to note that in our context, follow-relationships are directional, and following more users from the majority group does not necessarily mean having more followers from that group.
Instead, the observed advantage can be traced back to the real-graph recommendation policy which draws on common outgoing and incoming edges as a measure of similarity between users. 
Since the majority of follow-relationships (89.89\%) are formed within groups rather than across groups, minority users who follow more majority users tend to have more connections in common with other majority users, resulting in more recommendations to the majority group.
The bottom row of Figure~\ref{fig:real_graph_homophilic_minority_work_rec_to_maj} illustrates the average recommendations of professional content created by the minority group for majority group users, highlighting the role of topic preferences. Although there is a positive correlation between professional content preference and recommendations ($\rho=0.08$, $p\text{-value}<0.05$), interest in the other two topics appears to be more evidently connected to professional content recommendations. Specifically, minorities with a stronger preference for the majority-related topic receive more recommendations for their professional content to the majority group ($\rho=0.14$, $p\text{-value}<0.05$). In contrast, a higher interest in the minority-specific topic tends to correlate with fewer recommendations of work content to the majority group ($\rho=-0.23$, $p\text{-value}<0.05$).


Overall, our simulation results indicate that minorities who are more similar to the majority group, in terms of expressed topic interests and social connections, gain enhanced cross-group visibility for their professional content. In this way, recommendation algorithms can be seen as creating an incentive for assimilation into the majority group. 


\subsection{Marginal topic content by majority diminishes visibility of minorities}

\begin{figure}
    \centering
    \includegraphics[scale=0.6]{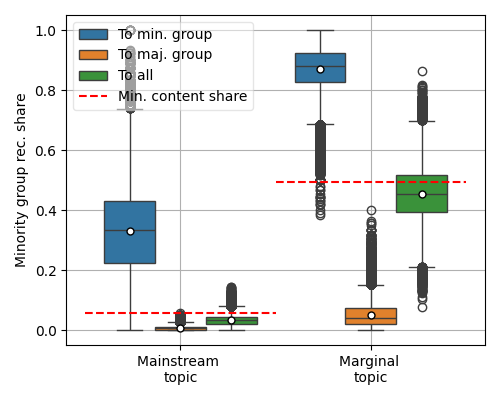}
    \caption{Average share of minority group recommendations compared across content topics for minority, majority, and all groups. Data covers 20 simulation runs with 10,000 time steps using \textbf{real graph recommendation policy and homophilic network structure}. Red lines depict minority group share in content creation for each group, white dots represent means.}
    \label{fig:real_graph_homophilic_box}
\end{figure}


So far, our attention has been directed towards recommending professional posts. These posts hold significance for two main reasons: First, they can directly contribute to tangible professional and career benefits. Second, they are equally appealing to users from both minority and majority groups. 
Besides those advantages, however, social media platforms also provide opportunities for collective identity formation and avenues for raising awareness, amplifying voices, and preserving cultural heritage \cite{brock2009you, jackson2020hashtagactivism, khazraee2018digitally}.
With this in mind, we turn our analysis towards recommendation of content with mainstream and marginal topics by both minority and majority group platform users. 

Figure~\ref{fig:real_graph_homophilic_box} illustrates the distribution of minority group shares among recommendations of mainstream and marginal topics, alongside the corresponding minority group shares of created content in the same category. In our simulation design, the minority group contributed only 5.72\% of mainstream topic content but accounted for 49.29\% of marginal topic content, influenced by both topic preferences and group sizes. Notably, minority group-created content is on average recommended less frequently as compared to majority content. Specifically, only 3.40\% of mainstream topic recommendations feature minority group posts, whereas 45.51\% of marginal topic recommendations incorporate minority group content. Further stratification by the population groups of recommendation receiving users reveals a strong tendency for in-group recommendations over cross-group recommendations which can be problematic for the aforementioned desire to increase visibility and amplify voices by the minority group. 
Despite the fact that minority group users create around half of the marginal topic content and are, on average more interested and engaged in the marginal topic than majority group users, only 4.95\% of marginal topic content recommendations to the majority group feature minority created content.

As a simplified illustrative example, consider an academic community consisting of a majority of native-born citizens and a minority group of immigrants. Suppose a topic of high relevance within the immigrant community centers around issues like navigating immigration processes, the economic and social dimensions of immigration, or challenges in cultural adaptation. Our simulations suggest that, under homophilic network dynamics and recommendation algorithms, majority users are more likely to encounter stories about the immigrant experience created by fellow native-born users. This occurs despite the firsthand insights offered by immigrants themselves, whose voices remain underrepresented in these recommendations.\footnote{As another example, consider a social platform used by an interdisciplinary group of academics with a shared interest in AI ethics, where the majority of users are from STEM fields (e.g., engineering, computer science) and a minority group consists of scholars from the humanities (e.g., philosophy and linguistics). Now, assume a topic of high relevance to the humanities scholars is issues about philosophy of mind (e.g., meaning, representation, consciousness). In this setting, our simulation findings suggest that, given a homophilic network structure and a recommendation policy based on previous interactions and common connections, STEM users are considerably more likely to receive recommendations on philosophy of mind content created by fellow STEM researchers (e.g., on whether large language models understand meanings), despite a significant portion of such discussions being generated by humanities scholars who specialize in the topic.}


On the flip side, when marginal topic recommendations are made to the minority group, minority created content has an increased share of 87.10\% which is higher than the minority group share of created content. 
Overall, while recommendation algorithms can facilitate community building among minority group users and increase visibility for \emph{marginal topics}, this does not necessarily translate to an amplification of minority voices and visibility for \emph{minority users}.

\section{Discussion}\label{sec:discussion}
Our findings highlight how specific components of widely used social media recommendation algorithms can lead to patterns of invisibility and homogenization. These patterns emerge, even though individual interest in and engagement with professional content is \textit{independent} of group identity. In this section, we first characterize the types of harms these patterns pose to individuals, groups, and epistemic communities. We then explore how this characterization extends existing philosophical understandings of the sources of epistemic harms, and consider the implications for science policy proposals. Finally, we outline potential strategies for addressing these harms and suggest directions for future research.


\subsection{Harms to individuals, groups, and epistemic communities}
Our results demonstrate how the social identity of researchers can significantly shape their visibility and opportunities in today's algorithmically-driven knowledge economies. Specifically, we show how, in the presence of seemingly neutral homophily-based communication dynamics, recommendation algorithms can disproportionately harm researchers from underrepresented or minority groups.

First, by rendering professional content created by minority researchers less visible, these algorithms undermine their status as knowers. In doing so, they effectively silence minority voices on professional topics and exclude them from broader information exchanges. These harms align with the harms of epistemic injustice \cite{fricker2007epistemic,mckinnon2016epistemic} and epistemic exclusion \cite{dotson2014conceptualizing,settles2021epistemic}, and can be seen as forms of such epistemic harms.

Second, our findings show how posts by minority users who resemble the majority---that is, those with similar preferences to the majority or more deeply embedded in majority network---can gain more majority group recommendations. In this way, these algorithms can critically shape \textit{expectations} about what ``successful minority researcher'' looks like. In this way, our findings echo similar patterns to what \citet{cheng2023social} refer to as \textit{social norm bias}, which arise due to the association between algorithmic predictions (e.g., classification of professional biographies) and individuals' adherence to inferred social norms (e.g. how researchers in some discipline present). While they focus on the harm of (supposedly fair) classification algorithms for individuals ``deviating from norms associated with a majority'', our results additionally highlight the risk posed by certain social media recommendation algorithms to individuals adhering to preferences associated with the minority group. 

Second, our findings reveal how posts by minority researchers who resemble the majority---whether through similar preferences or by being more embedded within majority networks---are more likely to be recommended to the majority group. This dynamic can shape harmful expectations about what a ``successful minority researcher'' should look like. In this sense, our findings echo what \citet{cheng2023social} describe as social norm bias: the association between algorithmic predictions (e.g., correct classification of professional biographies) and individuals' adherence to inferred social norms (e.g., the prevalent self-presentation of researchers in a given discipline). While Cheng et al. focus on algorithms disadvantaging those who \textit{deviate} from majority norms, our findings highlight a further risk posed by recommendation algorithms to minority researchers who \textit{maintain} preferences and associations aligned with their own communities.

In this way, these results underscore the risk that recommendation algorithms create incentives for \textit{assimilation}. Notably, even without direct knowledge of algorithm design, social media users may still feel these incentives. Research on \textit{algorithmic folk theories} shows how users form beliefs about the workings of algorithms (based on experience, rumors, or observations) and adjust their self-presentation and behavior accordingly~\cite{devito2018people,mayworm2024content}. For instance, Karizat et al. refer to \textit{the identity strainer theory}, where users believe that ``an algorithm filters content based on social identity, resulting in the suppression of marginalized social identities on a platform's social feed''~\cite[][p. 305]{karizat2021algorithmic}. Our findings offer empirical support for this folk theory, demonstrating that such outcomes can emerge from components of widely used content recommendation algorithms.

Finally, professional recommendation algorithms can harm minority group members even outside the professional topic. These algorithms diminish minority voices, even on topics where they may possess unique expertise (e.g., the immigrant experience). This not only introduces further epistemic harms but also causes representational harms~\cite{chien2024beyond} by perpetuating content that, despite pertaining to minority concerns, distorts or discounts minority perspectives. Such patterns could result in concerns that are similar to epistemic appropriation, where mainstream discussions about topics that largely emerge from and concern minority groups are largely detached from individuals from those groups~\cite{davis2018epistemic}. Such distortions arguably not only harms individuals from the minority group, but also the majority who are deprived of those users' insights. 

Beyond these harms to minority groups, the reduced opportunities for cross-group engagement pose risks to the epistemic community as a whole. Cross-group exchange plays a crucial role in realizing the epistemic and social benefits of diversity. Cognitive diversity---diversity of perspectives, lived experiences, methods, and ways of thinking---is often valued for its ability to improve the epistemic outcomes of the entire community~\cite{page2019diversity,rolin2019epistemic,steel2021information,muldoon2013diversity,wu2023epistemic}. Sociocultural diversity often correlates with cognitive diversity~\cite{page2019diversity}. As \citet{settles2021epistemic} note, for example, ``faculty of color are more likely than others
to have diverse approaches to their scholarship and to study populations and topics that do not fit neatly within these disciplinary norms'' (p. 495). By reducing the possibility of cross-group exchange, recommendation algorithms diminish cognitive diversity, and thus its benefits for the entire community.

This lack of cross-group exchange also undermines the social benefits of diversity and threatens social cohesion. Research shows that inter-group friendships can reduce prejudice and inter-group anxiety~\cite{levin2003effects}, while segregation generated by homophily can reinforce group-based biases and unfairness~\cite{o2019origins}. In sum, social media algorithms not only harm individuals from underrepresented groups but also negatively affect the broader epistemic communities. Understanding and addressing these harms will thus benefit not only minority groups, but also the epistemic community as a whole.

\subsection{Sources of epistemic harms, and science policy proposals}
These findings expand our understanding of the sources of epistemic harms, such as epistemic injustice and exclusion, experienced by researchers from minority groups. Typically, works on these topics have located the source of these harms at the \textit{psychological} level, such as in audiences’ (implicit or explicit) prejudice and bias~\cite{mckinnon2016epistemic,settles2021epistemic,fricker2007epistemic}. For instance, researchers have discussed biases that stem from harmful identity stereotypes (e.g., when members of minoritized groups are perceived as less credible or deserving) or from devaluing certain topics or methodologies (e.g., when socially-oriented applied research, often conducted by marginalized communities, are viewed as less rigorous). Recently, scholars have also begun identifying \textit{social} causes of these harms. For example, \citet{rubin2022structural} shows how unjustified disparities in citations can emerge due to structural features of epistemic networks---specifically, homophilic patterns of association---even in the absence of individual prejudice.

Our work extends this discussion by highlighting the distinct role of \textit{technology} as a potential source of systemic epistemic harms. As our findings demonstrate, recommendation algorithms can transform seemingly benign preferences for in-group non-work topics (e.g., discussions around culturally specific TV shows) into mechanisms of epistemic exclusion on professional topics. In this way, algorithms contribute to the invisibilization and exclusion of minority voices within epistemic communities. Critically, these harms are \textit{algorithmically generated} and are absent from traditional organizational settings that motivate (and are analogous to) our simulations, where differential engagement tends to be confined to non-work-related topics.\footnote{This technological source of harm interacts dynamically with psychological and social factors. For instance, the harmful effects of recommendation algorithms are particularly pronounced in homophilic networks. These effects, in turn, lead to a range of harms, including reduced cross-group engagement, which can undermine social cohesion and exacerbate prejudice. Thus, the interaction between technological, psychological, and social elements must be understood as a mutually reinforcing system~\cite[see ][]{liao2021oppressive}.}

This expanded understanding has important practical implications, particularly for science policy proposals that look to technology as a solution to persistent social problems. As mentioned earlier, social media platforms are often seen as tools for promoting openness and inclusion in epistemic communities. Some have even advocated for using alternative metrics based on social media engagement in academic promotion and tenure decisions~\citep{cabrera2018social,piwowar2013value}. However, our findings suggest that, without careful consideration of how recommendation algorithms function, such policies risk \textit{worsening} the very patterns of exclusion they aim to address. 

Our work aligns with recent philosophical calls for caution against proposals that seek to mitigate the negative effects of traditional gate-keeping mechanisms, such as peer review, by optimistically relying on technological alternatives (e.g., citation counts or search algorithm rankings)~\cite{rubin2022structural,desmond2024gatekeeping}. 
Although the workings of algorithms may be less visible than traditionally considered psychological, organizational, or social barriers to inclusion and openness, our results highlight that their impact on shaping information exchange is equally critical in our societies.

\subsection{Counteracting harms}
How might the harms discussed above be mitigated? Below, we consider several individual, technological, and policy strategies and interventions.\footnote{Note that these are not meant to be exhaustive.} 

The homophily-driven patterns of disclosure and engagement that inspire our simulations have been widely studied in traditional organizational settings. \citet{phillips2018diversity} suggest several strategies for combating their adverse impacts in such contexts. One individual-level strategy involves encouraging minorities to adjust their behavior by ``beginning their self-disclosure by sharing status-disconfirming interests that help them connect with others''~\cite[][p. 6]{phillips2018diversity}. In terms of our simulations, this aligns with our finding that minority individuals who more closely resemble the majority gain greater visibility. However, while \citet{phillips2018diversity} emphasize that this approach is not meant to promote assimilation or inauthenticity, but rather to bridge boundaries, it is easy to see the risks. This is particularly so, because the strategy implicitly places the burden of boundary-crossing on individuals with minority or underrepresented identities.

Another individual-level strategy, particularly suited for digital platforms, involves addressing the issue of context collapse, and the merging of the different audiences, topics, and norms. One approach might be to use platform affordances to separate audiences (e.g., creating distinct groups) or to create separate user profiles for different topics and audiences. While this strategy can mitigate some risks, it has limitations. First, as highlighted in organizational literature, opting out of non-professional conversations has its costs, since interactions on those topics often strengthen interpersonal relationships, enhancing one's sense of belonging and providing future professional opportunities. Moreover, making clear distinctions between audiences and topics is not always straightforward. 

The literature on algorithmic fairness in recommender systems offers various technical interventions that can help reduce the burden on individuals (particularly those from minority groups). Many of these methods intervene at the level of individual recommendation lists, such as requiring equal exposure across groups~\cite{Singh2018,Zehlike2017,Zehlike2020}. However, enforcing fairness of exposure~\cite{Singh2018,Zehlike2020} or equity of attention~\cite{Biega2018} can be misleading, as increased exposure does not necessarily result in higher utility for the user. Recent research emphasizes the importance of adopting contextual, long-term perspectives on recommendation fairness~\cite{Patro2022,Akpinar2022}. 


Finally, we can consider policy-level interventions. One such intervention is \textit{transparency}~\cite{nyhan2022black}. While transparency alone does not resolve the issues discussed here, it is critical for \textit{verifying} and \textit{understanding} the problem. As mentioned earlier, already social media users from minority and marginalized backgrounds \textit{perceive} that recommendation algorithms reduce the visibility of their content because of their identity~\cite{delmonaco2024you,karizat2021algorithmic}. Opacity about the true workings of algorithms---whether due to technical complexity or corporate secrecy---makes it difficult, if not impossible, for these users to verify their concerns and devise appropriate responses~\cite{thach2024visible}.\footnote{For example, \citet{delmonaco2024you} discuss how users from marginalized groups collaboratively develop and test their algorithmic folk theories about shadowbanning.}

The type of evidential support provided by our findings, made possible through insights into the algorithm's workings, can empower users and support their calls for accountability. Additional policy interventions could further our understanding of these harms and the development of contextually appropriate responses by securing and supporting meaningful research partnerships between academia and industry.

\subsection{Limitations and future directions}
While agent-based simulations offer a valuable tool for studying long-term recommendation dynamics, these methods also require a host of modeling assumptions about idealizations, operationalization, parameterization, and more~\cite{Patro2022,Friedler2021}. We have aimed to explain and justify these assumptions throughout the paper, yet our findings remain constrained by a number of simplifying assumptions.

For example, our simulation model is static in terms of both network structure and agent preferences and behavior. First, the composition of the community and the interconnection among its members remain fixed---no agents enter or leave the network, or follow or unfollow others. Second, individuals do not adapt their behavior in response to their perception of how the algorithms function. This overlooks the role of users as strategic agents, whose interests and behavior evolve over time~\cite{Tennenholtz2019,cotter2019playing,delmonaco2024you,karizat2021algorithmic,Goga2015,chakraborty2018equality}. As noted earlier, this simplification was motivated by our interest in isolating the effects of the recommendation algorithm. Future research could examine the potential impacts of various strategies adopted by users, particularly those with minority identities, in response to perceived algorithmic suppression~\cite[see][]{karizat2021algorithmic,delmonaco2024you}.

Our analysis has focused on content recommendation from In-Network Sources, specifically examining the impact of deploying algorithms similar to RealGraph. However, social media platforms also recommend content from Out-of-Network Sources---those not followed by users. Although the specifics of these systems are beyond the scope of this paper, the types of questions they aim to answer raise important concerns: ``What Tweets did the people I follow recently engage with?'', ``Who likes similar Tweets to me, and what else have they recently liked?'' or ``What Tweets and Users are similar to my interests?''~\cite{twitter}. It is not difficult to see how recommendations based on such queries could \textit{amplify} the issues we identified in this study. Future research should explore this hypothesis by investigating in more detail the components of systems designed to facilitate such recommendations, such as GraphJet and SimClusters~\cite{satuluri2020simclusters,sharma2016graphjet,twitter}.

More broadly, we hope this work encourages deeper exploration of the impact of AI and data-driven technologies on epistemic communities. These tools now influence nearly every aspect of these communities---from college admissions and tutoring, to scholarly search engines and data analysis, to peer review and grant management, to digital communication and networking~\cite{zhai2021review,messeri2024artificial}. Our study takes an initial step toward highlighting the significant consequences of these tools, particularly the often hidden social power embedded in their design choices.

\bibliographystyle{ACM-Reference-Format}
\bibliography{refs}


\end{document}